\documentclass{iaus}
\usepackage{graphicx}

\title[~~Mars simulations and bacterial life] 
{New approaches to the exploration: \\ planet Mars and bacterial life.}

\author[Galletta, Bertoloni \& D'Alessandro]   
{Giuseppe Galletta$^{1,2}$, Giulio Bertoloni$^3$  \and Maurizio D'Alessandro$^4$ }

\affiliation{$^1$Dipartimento di Astronomia, Universit\`a di Padova, vicolo Osservatorio 3, \\
35122 Padova, Italy \\ 
$^2$CISAS "G. Colombo", Universit\`a di Padova, via Venezia 15, 35131 Padova, Italy \\
email: {\tt giuseppe.galletta@unipd.it} \\[\affilskip]
$^3$Dipartimento di Istologia, Microbiologia e Biotecnologie Mediche, Universit\`a di Padova,  via Gabelli, 35121 Padova, Italy \\ email: {\tt giulio.bertoloni@unipd.it} \\[\affilskip]
$^4$INAF - Osservatorio Astronomico di Padova,  vicolo Osservatorio 5, 35122 Padova, Italy \\ email: {\tt maurizio.dalessandro@oapd.inaf.it} }

\pubyear{2010}
\volume{269}  
\pagerange{001--005}
\jname{Galileo's Medicean Moons – their impact on 400 years of Discovery}
\editors{C. Barbieri, S. Chakrabarti, M. Coradini \& M. Lazzarin, eds.}
\begin{document}

\maketitle

\begin{abstract}
Planet Mars' past environmental conditions were similar to the early Earth, but nowadays they are similar to those of a very cold desert, irradiated by intense solar UV light. However, some terrestrial lifeform showed the capability to adapt to very harsh environments, similar to the extreme condition of the Red Planet. In addition, recent discoveries of water in the Martian permafrost and of methane in the Martian atmosphere, have generated optimism regarding a potentially active subsurface Mars' biosphere. These findings increase the possibility of finding traces of life on a planet like Mars.

However, before landing on Mars with dedicated biological experiments, it is necessary to understand the possibilities of finding life in the present Martian conditions. Finding a lifeform able to survive in Martian environment conditions may have a double meaning: increasing the hope of discovering extraterrestrial life and defining the limits for a terrestrial contamination of planet Mars.

In this paper we present the Martian environment simulators LISA and mini-LISA, operating at the Astronomical Observatory of Padua, Italy. They have been designed to simulate the conditions on the surface of planet Mars (atmospheric pressure,0.6-0.9 kPa; temperature from -120 to 20 $^\circ$C, Martian-like atmospheric composition and UV radiation). In particular, we describe the  mini-LISA simulator, that allows to perform experiments with no time limits, by weekly refueling the liquid nitrogen reservoir.  

Various kind of experiments may be performed in the simulators, from inorganic chemistry to biological activity. They are offered as experimental facilities to groups interested in studying the processes that happen on the Martian surface or under its dust cover.
  
\end{abstract}

\firstsection 
\section{Introduction}

Planet Mars represents a challenge to several theories about the origin of life on Earth. If life on Earth appeared soon after the formation of the oceans, about 3.8 Gyr ago (\cite{pascale,westall}), similar conditions may hold in the planet Mars, where billions years ago a volcanic activity was probably producing a greenhouse effect able to keep atmospheric pressure high enough to maintain liquid water on the surface. 

However, since Mars has  the 53\% of size of the Earth, it cooled more quickly. We know that volcanic activity on Mars terminated long times ago, leaving enormous quescient volcanoes whose flanks are eroded by wind and thermal variations. When the planet reached the cold and dry condition we can observe today, liquid water, that's so important for life as we know it, should have disappeared from its surface. The mechanisms that produced life on Earth are still under debate, but if some kind of lifeforms appeared on Mars in the past, similar or different from the terrestrial ones, some colony of living being may have adapted to climate changes and may have survived in some ecologic niches near the soil or deep under the  surface. 

This hypothesis is supported by the fact that terrestrial lifeforms show a strong capability to adapt to very harsh environments and to survive even for some period in place that is more inhospitable than Mars' surface: the circumterrestrial space (see for instance in \cite{sancho}). If a now dormant lifeform has adapted to Mars surface, despite the strong ultraviolet radiation of the Sun reaching the soil, we may expect to find trace of it in future explorations of the red planet. Digging under the Martian surface may bring more information about past or present lifeforms. Life may have harbored in the Martian underground, where  the diurnal and seasonal temperature changes are lower than those happening on the ground, the UV light is absorbed by the upper dust or ice/snow layers (see \cite{cockell2002}) and the pressure increase with the deepness reaching values at which water is in liquid state.  The recent discoveries of water in the Martian permafrost by means of orbiter's instruments (\cite{demidov}) and by excavation done with the robotic arm of Phoenix lander, along with the discovery of methane in the Martian atmosphere (\cite{krasnopolsky2004, formisano2004}) have generated optimism regarding a potentially active subsurface Mars biosphere. 

While waiting for new on site biological experiments, we may however understand what physical mechanisms in the Martian environment could be theratogens for lifeforms, through some experiments done in simulated Martian environmental conditions. Those environmental conditions are: temperature ranging from a maximum of $\sim$20 $^\circ$C in the tropical summer to -140 $^\circ$C in the polar winter; pressures around 7 mbar(=700 Pa); UV-C flux around 3.3 W/m$^2$ according to the estimates of \cite{patel2004}; atmospheric gases CO$_2$ 95.5\%, N$_2$ 	2.7\%, Ar 	1.6\%	 O$_2$ 0.13\%, CO 0.07\%.

In addition, while exploring Mars we can incidentally or voluntary drop terrestrial lifeforms, able to survive and to be reactivated once brought in laboratories. Finding conditions that allow the survival of lifeforms in a Martian environment may have then a double value: increasing the hope of finding extraterrestrial life and defining the limits for a terrestrial contamination of planet Mars.

\section{Mars simulations}

\subsection{LISA}

To test the survival of lifeforms on Mars, we projected and built two simulators of the Martian environments where to perform experiments on terrestrial bacteria strains. The bigger one is called LISA (Laboratorio Italiano Simulazione Ambienti), described in \cite{galletta2006, galletta2007}, that allows six simultaneous experiments in distinct reaction cells of 250 cc. To reach the low Martian temperatures, the internal plate hosting the reaction cells is cooled by an external Liquid Nitrogen closed-loop cooling system. This allow us to reach temperatures, inside the chamber, down to 150 K (-123 $^\circ$C). An electric resistance is able to rise the temperature to the environment one and may reach for cleaning and degassing purposes after the experiments +57 $^\circ$C. LISA has been used for many experiments (\cite{visentin2009}) but has a limit on the length of a single experiment. The large size of the camera, 70 cm in diameter, the massive plate that holds the six reaction cells, giving a good thermal inertia, are unfavorables for the liquid nitrogen consumption. With a 500 liter reservoir, we are able nowadays to perform experiments not longer than 30 hours ($\sim$1.25 Martian Sol). 

To overcome this problem and to extend the length of experiments to one week or one month, and by taking into account the experience gained in the past years in the field of imaging sensors cooling technique (Charge Coupled Device), we designed and built  MiniLISA, a smaller but more efficient, from the point of view of the Liquid Nitrogen use, version of LISA.

\begin{figure}
 \includegraphics[width=7cm]{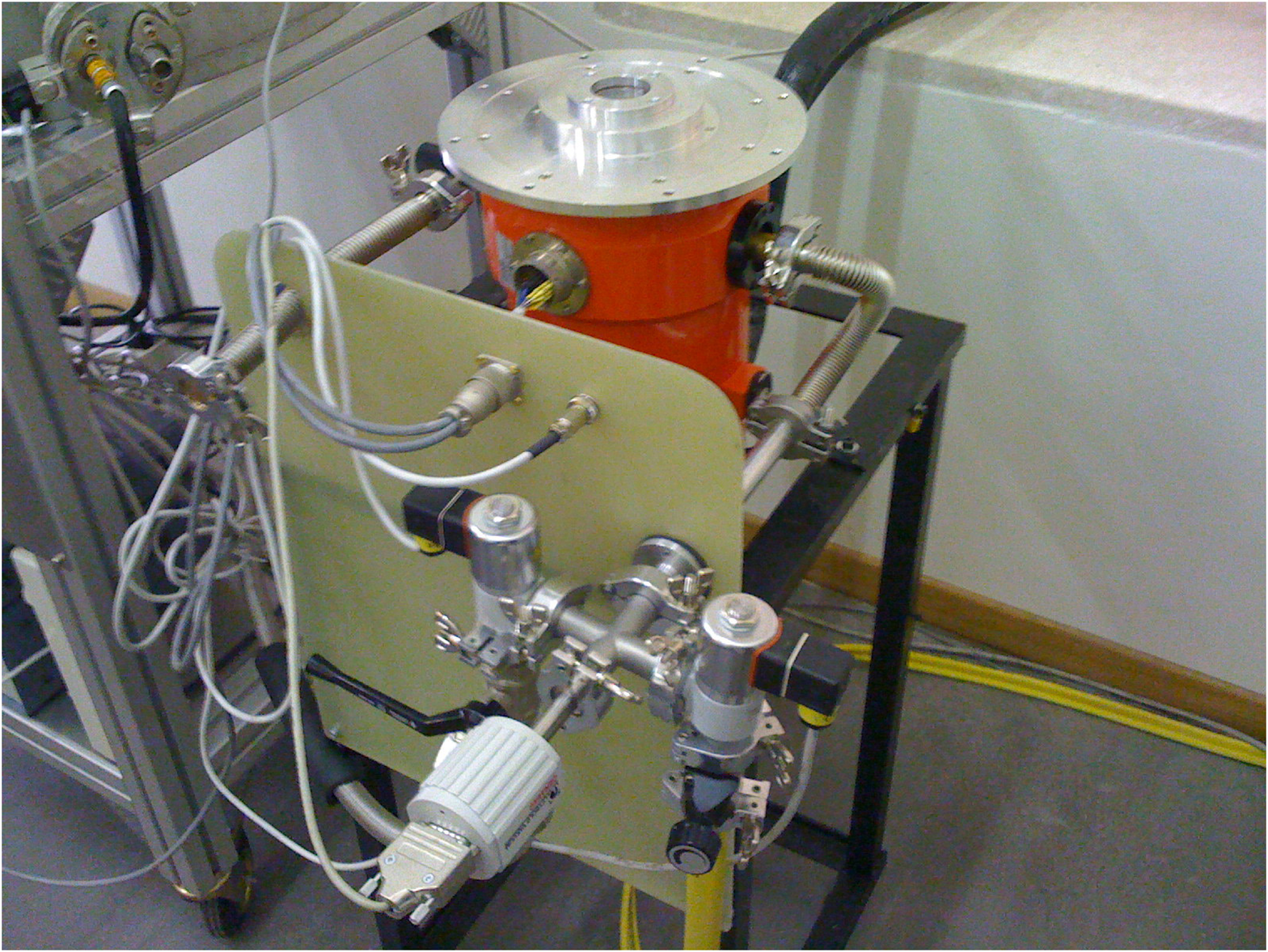}
 \includegraphics[width=7cm]{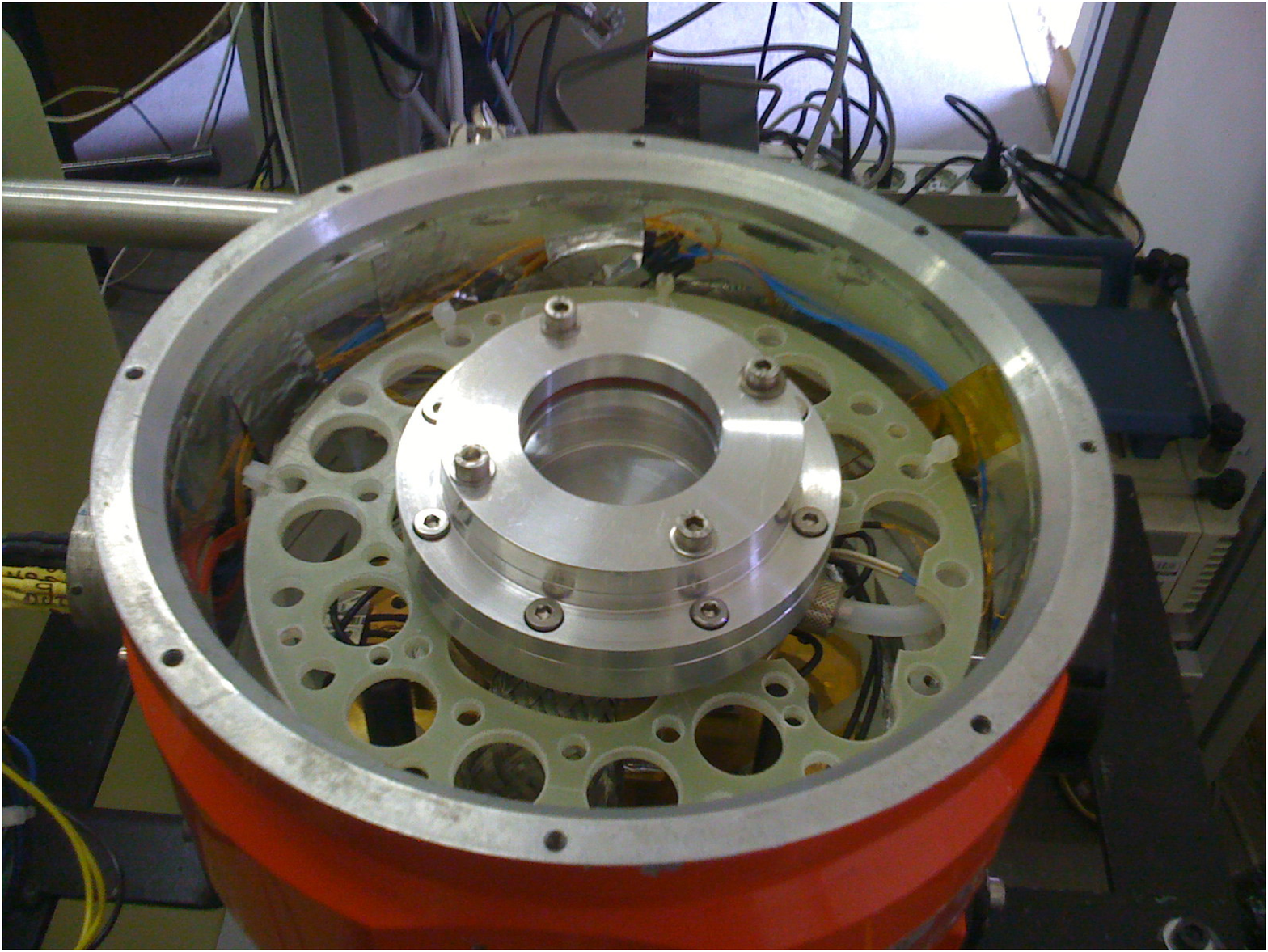} 
 \caption{The Mini-LISA simulator. \textit{Left}: The cryostat with the control cables and the atmospheric control system, with the two electro-valves regulating the gas flux and the pressure probe. \textit{Right}:The inner part of the simulator, with at the center the reaction cell, hosting the two aluminum Petri dishes containing the biological samples. }
   \label{miniLISA}
\end{figure}

\subsection{mini-LISA}

The idea behind this project was to use a standard commercial dewar (from Oxford Instruments) and to modify it for our applications. This dewar, normally, is used to cool image sensors (Charge Coupled Device) routinely used for imaging or spectroscopy at optical telescopes. As shown in Figure \ref{miniLISA} the reaction cell is isolated from the rest of the Dewar. This allow to evacuate the space between the cell and the dewar walls by the use of a vacuum turbo-pump, able to produce pressures down to 4-5 10$^{-3}$ Pa. The  air in the reaction cell, at a pressure of $\sim$100 kPa is evacuated and a mixture mainly composed by carbon dioxide is pumped inside. Then the mixture is evacuated using a control system and a second vacuum pump (see Figure \ref{miniLISA}, left panel) and refilled several time, every time at a lower pressure, in order to ``clean'' the reaction cell from terrestrial atmosphere residuals and to obtain a Martian-like atmosphere at a pressure of 0.7 kPa . 
The low pressure isolates the reaction cell from the external environment, minimizing the thermal exchange between the samples closed inside it and the laboratory environment. With this configuration and procedure, the 500 liter liquid nitrogen reservoir is able to produce 10 days of experiment or, being refueled once per week, an experiment not limited in time.  

The drawback of MiniLisa is that only one reaction cell can be installed inside the dewar. To allow the presence of a comparison sample in the same reaction cell, we used a special combination of aluminum dishes, similar to the classical Petri dishes for biological samples. The dishes are stacked up one upon the other and inserted in the bottom of the reaction cell. In this way the dish on the top is exposed to the UV radiation entering from the quartz (Suprasil) window, while the bottom one is protected from the light. Because of this configuration, the number of biological samples that can be used is limited by the surface of about 60 mm$^2$ of the aluminum dishes.  

\subsection{bio-packages}
In both simulators the biological samples are deposited in small glass plates of $\sim$1 cm hosted in the aluminum dishes inside the reaction cell. This procedure has been adopted to make easier the removal of the samples after every experiments. Aluminum dishes in fact, although flattened in the inner part, have microscopic irregularities on their surfaces that may retain some biological residuals contaminating the next experiments. The atmospheric gases transit in the cell by means of a pipe circuit and the exit of biological samples from the cells is prevented by a set of mechanic filter. The quartz windows may be closed through some aluminum covers that stop the light for a predetermined time, in order to irradiate the samples for a controlled time. Biological samples are then immersed in the gas at Martian pressure, temperature and irradiation without being in contact with the laboratory environment for the whole length of the experiments. 

Inside LISA we studied the survival of several bacterial strains belonging to the genus \textit{Deinococcu}s, and to the endospore forming genera \textit{Bacillus}.  The preliminary results of our studies are presented in \cite{galletta_frascati}.

In addition to the use for our experiments, LISA simulators are offered as experiment facilities to research groups willing to study soil samples, small instruments or lifeforms in Martian conditions. At present, some tests have been performed for biologists of the German Aerospace Center in Cologne and of the Department of Animal Biology at the University of Modena. 

\acknowledgements

The authors thanks the Air Liquide Italy for kindly supporting the LISA simulators with liquid nitrogen supplies. This work has been supported by the research funds (ex 60\%) of Padua University.




\end{document}